# INVESTIGATION ON THE SHOOTING METHOD ABILITY TO SOLVE DIFFERENT MOORING LINES BOUNDARY CONDITION TYPES

**Florian Surmont**
Research Department
Bureau Veritas
Saint Herblain, France

**Damien Coache**
Research Department
Bureau Veritas
Saint Herblain, France

**ABSTRACT**

The study of undersea cables and mooring lines statics remains an unavoidable subject of simulation in offshore field for either steady-state analysis or dynamic simulation initialization. Whether the study concerns mooring systems pinned both at seabed and floating platform, cables towed by a moving underwater system or when special links such as stiffeners are needed, the ability to model every combination is a key point.

To do so the authors propose to investigate the use of the shooting method to solve the two point boundary value problem (TPBVP) associated with Dirichlet, Robin or mixed boundary conditions representing respectively, displacement, force and force/displacement boundary conditions.

3D nonlinear static string calculations are confronted to a semi-analytic formulation established from the catenary closed form equations. The comparisons are performed on various pairs of boundary conditions developed in five configurations.

**NOMENCLATURE**
TPBVP Two Point Boundary Value Problem
IVP Initial Value Problem
ODE Ordinary Differential Equation
RKF45 Runge Kutta Fehlberg 4-5
BC Boundary Condition

**INTRODUCTION**

In this paper the mooring lines and cables are modeled as nonlinear extensible elastic strings handling only traction along their centerline. This particular model is indeed extensively used in offshore industry due to its numerical time efficiency compared to more detailed beams equations. Three main techniques are used to solve static string equations [1]: dynamic relaxation, discretization based methods (lumped mass, finite difference, finite elements) and shooting methods. The shooting method has been previously introduced in offshore field to analyze three-dimensional steady-state configuration of underwater flexible cable problem. Application cases were developed for the resolution of the line profile and tension under gravity and current loads on a single cable linking a surface vessel to a buggy at seabed, and single line towing configurations [2,3], where cables are pinned at both extremities. In some offshore commercial software, catenary equations are solved using a derivative of the shooting method based on the same procedure developed in the catenary algebraic equations paragraph of this article. Moreover, the present paper develops a shooting method equations set for full 3D static equations where external loads are arbitrary and not restricted to gravity alone. This differs from current commercial software approach by the fact that these equations could also be used in a current drag context as stated in previous author's work [4] where solution profile differs from a catenary shape. In different contexts than offshore engineering, the shooting method [5,6] is used in a wide range of application fields, from medical robotic assisted surgery [7,8] to periodic determination of stator-rotor assembly [9]. One of the first to investigate the use of optimal control theory and the Pontryagin Maximum Principle in solving TPBVP was Keller [10,11] by developing the single shooting method and multiple shooting method. Since then several shooting methods "types" can be found in the literature from the multiple shooting method [12,13], to continuation techniques [14], as well as shooting method applied on Lie group manifolds [15].

Based on previous authors' work [4] this paper aims to make use of a classical single shooting method approach to solve the statics of a string with a special emphasis on its capability to take into account several kinds of boundary conditions often present in offshore industry.

In a first step the general 3D TPBVP static equations based on [16] are presented with the introduction of arbitrary 3D external distributed loads. Particular attention is given to the description of the different boundary conditions types, namely Dirichlet, Robin and mixed boundary conditions. In the context of studying strings, these kinds of boundary conditions correspond to spherical, prismatic, planar joints, imposed forces at the tip or stiffeners which are often used in offshore field models.

Then the shooting method principle is presented for solving this kind of problem, where the TPBVP is solved iteratively as a succession of initial value problems (IVPs). The proposed motion equation state variables formalism, associated with the shooting method formalism let us write naturally the mentioned boundary conditions to be able to impose a pure force, a pure displacement or a mix force/displacement in the same way.

In a third step, the present paper investigates how the shooting method can solve the TPBVP of a slender body under the mentioned boundary conditions. Five validation cases are proposed corresponding to various boundary conditions combinations: ball-ball, ball-force, ball-prismatic, ball-spring, ball-prismatic/spring. Each case is compared to a semi-analytic formulation based on algebraic catenary closed form equations.

## TPBVP EQUATION

The TPBVP for static strings is constituted of a set of ordinary differential equations (ODEs) associated with two boundary conditions at segment extremities. String theory is derived from [16] formalism, based on the following assumptions:

(i) a configuration of the string is defined as a set of material points (sections) having the geometrical property to be a curve in $\mathbb{R}^3$,
(ii) the string is 'perfectly flexible', withstanding only pure stretch deformations (i.e. deformations only along the centerline)
(iii) geometrical and mechanical functions defining the string are sufficiently regular to be derived accordingly

### Geometrical deformation

From (i), a material point in space is described by the vector $r(s)$ identified in global coordinates system by a parameter $s$. This parameter is taken as the curvilinear abscissa of the reference configuration (zero acting force configuration) $r^0(s)$ such that $|\partial_s r^0(s)| = 1$, where $\partial_s \cdot$ denotes the partial derivation with regards to parameter $s$. From (i) and (ii) the stretch $v$ is defined as $v(s) = |\partial_s r(s)| > 0$. In the reference configuration, the curvilinear abscissa is defined on the interval $s \in [0, L]$. The unstretched length of the material segment is $L = \int_0^L v^0(s)ds$ whereas the stretch length is obtained by $l = \int_0^L v(s)ds$.

From the previous geometrical description, we introduce the stretch vector $v(s)$ such that:

$$\partial_s r(s) = v(s) = v(s)\frac{\partial_s r(s)}{|\partial_s r(s)|} \quad (1)$$

### Balance equation

The forces acting on a generic segment $(s_1, s_2)$ where $0 \leq s_1 < s_2 \leq L$ are defined as $n^+(s_2)$ the contact (internal) force exerted by segment $(s_2, L)$ on $(s_1, s_2)$ and $-n^-(s_1)$ the contact (internal) force exerted by segment $(0, s_1)$ on $(s_1, s_2)$. It is assumed that all other body (external) forces are of the form $\int_{s_1}^{s_2} f(\xi)d\xi$. $f(s)$ stands for arbitrary distributed forces per unit length acting on the string. Vectors $n^+$, $n^-$ and $f$ are defined in global coordinates.

From above definitions, the static balance equation on segment $(s_1, s_2)$ is [16]:

$$n^+(s_2) - n^-(s_1) + \int_{s_1}^{s_2} f(\xi)d\xi = 0 \quad (2)$$

Holding for all $0 < s_1 < s_2 < L$. The required continuity (iii) imposes for $s_1 \to s_2$: $n^+(s_2) = n^-(s_2) = n(s)$.

The sign convention adopted for $n$ is then: positive contact force when acting from a material point situated at $s + ds$ on a material point at $s$. Eqn. (2) then resolves to:

$$n(s) - n(s_1) = -\int_{s_1}^{s} f(\xi)d\xi \quad (3)$$

Finally by differentiating Eqn. (3) with respect to $s$ the local static equilibrium equation is obtained:

$$\partial_s n(s) = -f(s) \quad (4)$$

Equations (1) and (4) constitute the two differential equations of our problem. Note that these two equations are expressed in a *global* Cartesian system, meaning that the approach adopted follows a spatial - or Lagrangian - description.

Authors in [2,3] preferred global cylindrical coordinates to describe the internal tension $n$ in the shooting method. This approach works well when the system is pinned at both ends. As soon as other boundary conditions are considered, global cylindrical coordinates description for the internal tension leads to unnecessary complex expressions when processing partial imposed initial vector state in the shooting process. That is why Cartesian coordinates convention has been adopted to describe $n$.

Eqn. (4) stands for the continuous static balance equation, in which external distributed loads $f(s)$ as well as internal contact forces $n(s)$ are supposed continuous by virtue of (iii). Concentrated forces can be formally introduced in Eqn. (4) by making use of distributions [18] to obtain the balance equation $\partial_s n(s) = -f(s) - \sum_i \delta(s - s_i) F_i$ in which $\delta$ is the Dirac delta function. Doing so does not assume the resolution method used afterwards.

If the resolution method is known and compatible with discretization or segment "cuts", one can introduce external concentrated forces later in the resolution process. Let $s_i$ be the curvilinear abscissa at which the external concentrated force $F_i$ is applied. In finite elements or finite differences methods, a specific node would be introduced at $s_i$ and $F_i$ would be added to the external nodal force vector. In a shooting method resolution, one would use the multi-shooting technique to handle this kind of discontinuity. The principle is to "cut" the domain $[0, L]$ into subdomains such that $[0, L] = [0, s_1] \cup \cdots \cup [s_n, L]$, where Eqn. (4) stands and continuity properties of each field hold inside each subdomain $[s_i, s_{i+1}]$. The reader should refer to [4] for more information about the multi-shooting technique and its application to handle discontinuities and multi-segment configurations in an offshore context.

**Constitutive equation**

The constitutive law expresses the material properties providing the relation between the contact forces $n$ and the change of shape of the string for any configuration $r$. Assumption (ii) requires that $\forall s$:

$$\partial_s r(s) \times n(s) = 0 \qquad (5)$$

Substituting Eqn. (1) into Eqn. (5) yields the following definition for the constitutive law:

$$v(s) = \hat{v}(|n(s)|, s) \frac{n(s)}{|n(s)|} \qquad (6)$$

$\hat{v}(|n(s)|, s) = v(s) > 0$ is the stretch function previously introduced, which depends only on the norm of the tension vector. Note that this definition imposes $|n(s)| \neq 0$.
Moreover in the present paper, the string is limited to handle traction only.
Eqn. (6) represents the constitutive law of a string in a general form. By adjusting the function $\hat{v}(|n(s)|, s)$ one can model linear elastic, hyper-elastic or unstretchable behavior, see [4].
In the following we will assume an elastic behavior of the string such that the stretch function is:

$$\hat{v}(|n(s)|, s) = 1 + \frac{|n(s)|}{EA} \qquad (7)$$

Eqn. (6) together with (7) form the constitutive equations of the general case and its specialization to a linear elastic material such that Eqn. (8) holds.

$$v(s) = \left(1 + \frac{|n(s)|}{EA}\right) \frac{n(s)}{|n(s)|} \qquad (8)$$

Finally, substituting Eqn. (8) in Eqn. (1) along with Eqn. (4) yields the following ODE system:

$$\begin{cases} \partial_s r(s) = \left(1 + \frac{|n(s)|}{EA}\right) \frac{n(s)}{|n(s)|} \\ \partial_s n(s) = -f(s) \end{cases} \qquad (9)$$

**Boundary conditions**

Boundary conditions are mathematical definitions of specific relations between variables of a problem on the boundaries of a specific domain. There are four main boundary conditions types: Cauchy, Dirichlet, Neumann and Robin boundary conditions. A Cauchy BC fully defines the state vector and its derivative for one value of the state parameter (curvilinear abscissa, time, etc…) and thus defines an IVPs. In case of a TPBVP, the other three types describe BCs defined at both ends of the problem. The Dirichlet BC defines the state vector, while the Neumann BC states its derivative at both ends. The Robin BC defines the state vector at one end, and its derivative at the other. Finally one can make use of a mix of above BCs.

Within the cable formalism, the mathematical boundary conditions can be brought closer to the physical joints assemblies used in offshore industry through the background of mechanical kinematic joints (linkages). Indeed, in the case of strings the domain studied is parameterized by a one dimensional parameter, the curvilinear abscissa $s$, belonging to the interval $[0, L]$. The boundaries of this interval are then reduced to the material points at $s = 0$ and $s = L$. At these interfaces one can *connect* a kinematic joint as described further. A key point is the duality between twists (kinematic screw) and wrenches (force and momentum screw) of such linkage. In this paper we consider the physical joints as perfectly frictionless mechanical linkages to impose the boundary conditions. That is to say the power of the linkage's internal contact forces is considered to be null. The power of the contact forces is computed through the screw scalar product of the twist and wrench. Then, when a kinematic degree of freedom is unconstrained, the corresponding velocity is null, but the contact force is undetermined. In this case we will speak about constrained static degree of freedom. On the contrary, when a motion is kinematically constrained, it is statically free with null corresponding force or moment components. In the case of strings kinematic assumption (ii), all rotations are free to evolve and only translational motions can be constrained. Table 1 summarizes the four kinematic constraints compatible with strings geometrical parameterization. It also illustrates the duality between the linear velocities $p_x, p_y, p_z$ (twist moment) and the forces $X, Y, Z$ (wrench resultant), along with the rotational velocities $\omega_x, \omega_y, \omega_z$ (twist resultant) and the torques (wrench moment) which are null in the case of strings.

## TABLE 1. STRING COMPATIBLE LINKAGES

| | Joint type | Reduction point $M$ | Admissible twists | Admissible wrenches |
|---|---|---|---|---|
| 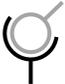 | **Spherical** of center $A$ | Point $A$ | $\begin{Bmatrix} \omega_x & 0 \\ \omega_y & 0 \\ \omega_z & 0 \end{Bmatrix}_A$ | $\begin{Bmatrix} X & 0 \\ Y & 0 \\ Z & 0 \end{Bmatrix}_A$ |
| 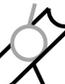 | **Linear annular** of center $A$, and axis $x$ | Point $A$ | $\begin{Bmatrix} \omega_x & p_x \\ \omega_y & 0 \\ \omega_z & 0 \end{Bmatrix}_A$ | $\begin{Bmatrix} 0 & 0 \\ Y & 0 \\ Z & 0 \end{Bmatrix}_A$ |
| 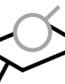 | **Punctual** of center $A$, and normal $z$ | Along normal $z$ | $\begin{Bmatrix} \omega_x & p_x \\ \omega_y & p_y \\ \omega_z & 0 \end{Bmatrix}_M$ | $\begin{Bmatrix} 0 & 0 \\ 0 & 0 \\ Z & 0 \end{Bmatrix}_M$ |
| 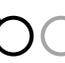 | **Free** | Every point in space | $\begin{Bmatrix} \omega_x & p_x \\ \omega_y & p_y \\ \omega_z & p_z \end{Bmatrix}_M$ | $\begin{Bmatrix} 0 & 0 \\ 0 & 0 \\ 0 & 0 \end{Bmatrix}_M$ |

In static studies, since no velocities are taken into account, the relative twist cannot be used directly and integrated positions and orientations are used instead. The global geometrical parameterization of the string is then retrieved from relative motions through vector composition. Because of (i) and (ii) only the translational part is considered. Hence there is a direct correspondence between $r$ and $n$ through the considered joint. Per the previous, a clamp link is equivalent to a spherical joint, prismatic and cylindrical joints are equivalent to a linear annular linkage as well as a planar joint is equivalent to a punctual linkage.

In order to be able to model imposed forces or springs, previous linkages definitions can be extended by replacing all zero forces components by functions of the motion. Doing so, the mechanical power of the contact forces is still considered null and this is equivalent to adding another specific wrench - for the imposed force or the spring - to the joint's internal force screw. If $n_{0,L}$ stands for the boundary value of the internal force at $s = 0$ or $s = L$, admissible statics presented in Table 2 is used. This table gives the different boundary conditions studied in the validation section. The punctual boundary condition has been omitted deliberately because catenary solutions, where the segment lies in a plane, will be investigated. In such conditions a punctual boundary condition is equivalent to a linear annular joint.

## TABLE 2. BCS FORCES COMPONENTS

| Joint type | Admissible statics |
|---|---|
| Spherical | $\mathbf{n_{0,L}} = [n_x \quad n_y \quad n_z]^T$ |
| Linear annular | $\mathbf{n_{0,L}} = [n_{x_{0,L}}(r) \quad n_y \quad n_z]^T$ |
| Punctual | $\mathbf{n_{0,L}} = [n_{x_{0,L}}(r) \quad n_{y_{0,L}}(r) \quad n_z]^T$ |
| Free | $\mathbf{n_{0,L}}(r) = [n_{x_{0,L}}(r) \quad n_{y_{0,L}}(r) \quad n_{z_{0,L}}(r)]^T$ |

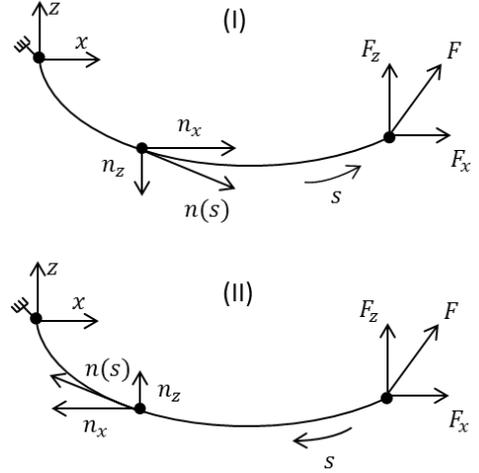

**FIGURE 1. SHOOTING BC FORCE SIGN CONVENTION. S GROWING POSITIVE ALONG $x$ (I) UPWARD, (II) DOWNWARD**

Forces functions $n_{x_0}(r)$, $n_{y_0}(r)$ and $n_{z_0}(r)$ can be chosen to match the required model such that, for a constant force Eqn. (10) applies, while a spring joint is achieved through Eqn. (11).

$$n_{i_{0,L}}(r) = \pm F_i \tag{10}$$

$$n_{i_{0,L}}(r) = \pm k(p - r) \cdot i \tag{11}$$

Vector $p$ is the constant coordinates of the reference point $P$ of the spring and vector $i$ stands either for $x$, $y$ or $z$ axis.
Hence, conceptually a spring is geometrically parameterized the same way as a free joint.

As shown on Fig. 1, the sign of the forces functions may be adjusted to $-1$ or $+1$ respectively at $s = 0$ and $s = L$, as requested by the chosen positive force acting from a material point at $s + ds$ on a material point at $s$.

Finally, considering the state vector in Eqn. (12) together with Eqn. (4) and (9), the TPBVP is written as follows:

$$\boldsymbol{\varphi}(s) = [\boldsymbol{r}(s), \boldsymbol{n}(s)]^T \tag{12}$$

$$\partial_s \boldsymbol{\varphi}(s) = \left[\left(1 + \frac{|\boldsymbol{n}(s)|}{EA}\right)\frac{\boldsymbol{n}(s)}{|\boldsymbol{n}(s)|}, \quad -\boldsymbol{f}(s)\right]^T \tag{13}$$

$$\overline{\boldsymbol{\varphi}}(s = 0) - \overline{\boldsymbol{\varphi}}_0 = \mathbf{0} \tag{14}$$

$$\overline{\boldsymbol{\varphi}}(s = L) - \overline{\boldsymbol{\varphi}}_L = \mathbf{0} \tag{15}$$

Notation $\overline{\cdot}$ represents the restriction to the constrained part of the state vector of Eqn. (12). $\overline{\boldsymbol{\varphi}}(s = 0, L)$ is the state vector evaluated at $s = 0$ or $s = L$ through integration of Eqn. (13) and restricted to the constrained part of the corresponding boundary condition. $\overline{\varphi}_{0,L}$ are the imposed values of the boundary condition.

Table 3 summarizes the constrained parts of each joint type. The linear annular joint is supposed directed along $x$ axis. Therefore, a spherical joint then corresponds to a Dirichlet BC, while free and pure spring joints are to be related to a Neumann BC. On the contrary, the (spring-)linear annular is a mixed BC.

**TABLE 3. BCS UNKNOWNS AND CONSTRAINTS**

|     | Joint type           | Unknown $\widetilde{\varphi}_{0,L}$ | | | Constrained $\overline{\varphi}_{0,L}$ | | |
| --- | -------------------- | --- | --- | --- | --- | --- | --- |
| (a) | Spherical            | $n_x$ | $n_y$ | $n_z$ | $x$ | $y$ | $z$ |
| (b) | Imposed Force        | $x$ | $y$ | $z$ | $n_x$ | $n_y$ | $n_z$ |
| (c) | Spring               | $x$ | $y$ | $z$ | $n_x$ | $n_y$ | $n_z$ |
| (d) | Linear annular       | $x$ | $n_y$ | $n_z$ | $n_x$ | $y$ | $z$ |
| (e) | Spring-linear annular | $x$ | $n_y$ | $n_z$ | $n_x$ | $y$ | $z$ |

## SHOOTING METHOD

The shooting method consists in transforming the TPBVP of Eqns. (13), (14), (15) into the search of the roots of a nonlinear constraint function being evaluated with a succession of IVPs.

The shooting problem is written as follows:

$$Root\ finding: \begin{cases} IVP: \begin{cases} \partial_s \varphi(s) = F(\varphi, s) \\ \varphi(s=0) = \varphi_0 \end{cases} \\ C(\varphi_0, \varphi(\varphi_0, L)) = \mathbf{0} \end{cases} \quad (16)$$

In this case, $C$ is a constraint vector function. $\varphi_0$ can be fully or partially unknown. The unknown part of $\varphi_0$ is noted $\widetilde{\varphi}_0$ and constitutes the unknowns of the root finding algorithm. The shooting method principle then consists in iterating over the guessed (unknown) starting missing part of the IVP in order to fulfill the constraint function $C$ evaluated through successive integrations to evaluate $\overline{\varphi}(s = L)$. The resolution process is described in Fig. 2.

In this paper, the continuous IVP equations are integrated with an embedded Runge-Kutta-Fehlberg 4-5 (RKF45) scheme. The constraint equation roots are sought by means of a Newton-Raphson algorithm, where the gradient is computed through finite differences. In the following sections, notation ~ will describe the unknowns of the Newton algorithm.

Concerning the use of an adaptive integrator, the RKF45 absolute error parameter corresponds to the cumulated error between the two schemes. It follows that adapting integration steps all over the line, leads to a cumulated absolute error between a RK4 and a RK5 scheme never exceeding prescribed integration absolute tolerance, which let us be confident about the precision of integration at line's end, see [4]. Hence the number of integration points is kept to a minimum while reaching the required accuracy.

For the case of strings and in order to reduce the number of unknowns in the Newton root finding method, we suppose half of the state vector of Eqn. (12) known (constrained) at $s = 0$ such that $\varphi_0 = \varphi(s = 0) = [\overline{\varphi}_0 \quad \widetilde{\varphi}_0]^T$. Both constrained and unknown components of the state vector $\varphi(s)$ are detailed in Table 3. The single shooting problem is given in Eqn. (17).

$$\begin{cases} \partial_s \varphi(s) = \left[\left(1 + \frac{|n(s)|}{EA}\right) \frac{n(s)}{|n(s)|}, \quad -f(s)\right]^T \\ \varphi_0 = [\overline{\varphi}_0 \quad \widetilde{\varphi}_0]^T \\ C(\varphi_0, \varphi(\varphi_0, L)) = \overline{\varphi}(s = L) - \overline{\varphi}_L \end{cases} \quad (17)$$

Note that within this formulation $\widetilde{\varphi}_0$ and $C$ have the same size of only 3. Moreover, in the present description, the unknown part $\widetilde{\varphi}_0$ of $\varphi_0$ depends on the boundary condition chosen at $s = 0$. Table 3 summarizes the known and unknown parts of each joint type. It should be highlighted that although all geometrical nonlinearities are taken into account the Jacobian of this problem is only a 3x3 matrix. Inverting a 3x3 matrix is then very effective compared to nonlinear discretized approaches where the Jacobian size depends on the squared number of elements.

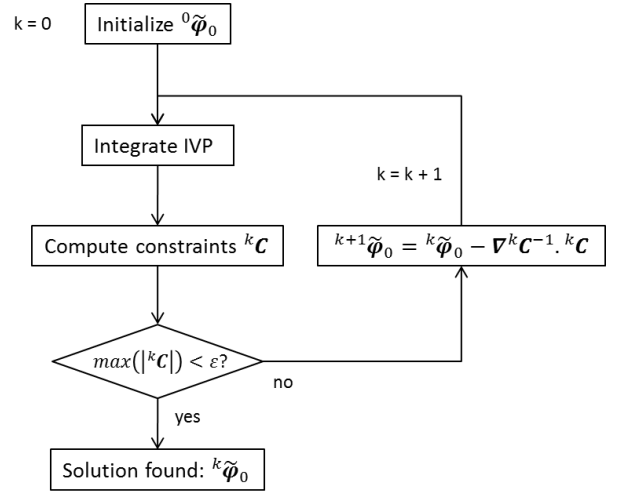

**FIGURE 2. SHOOTING METHOD RESOLUTION PROCESS**

## REFERENCE SOLUTION

In this part, a reference solution based on catenary solution is presented. Catenary cases are studied here in order to have a semi-analytical solution to be compared to the numerical results. Note that the shooting method proposed above has a wider range of application than catenary solutions only. Five test cases are considered, corresponding to various boundary conditions. Left end will always be a spherical (or ball) joint. On the right end side, the following five BCs will be considered: (a) spherical, (b) imposed force, (c) spring only, (d) linear annular, (e) spring-linear annular. In addition, the solution cannot depend on the choice of the curvilinear abscissa origin. Then all five cases are doubled in two groups where $s$ is growing positive along $\boldsymbol{x}$ (I) upward, or along $\boldsymbol{x}$ (II) downward. Therefore a total of 10 configurations are compared to semi-analytical results. To do so, first the catenary equations are developed. A single segment is considered, having uniform cross section characteristics. It is supposed fully stretchable; following elastic constitutive law of Eqn.(7). Then the semi-analytical resolution process is described. Special attention is given to the different BCs couples.

**Catenary algebraic equations**

The static in plane extensible hanging algebraic closed form for a single segment has been widely studied [17] and used in a number of commercial software; only the main results are reminded here. The configuration is described in Fig. 3. Suppose a line hanging in a fluid of volumetric mass $\rho_f$, of Young modulus $E$, sectional cross area $A$, volumic mass $\rho$ and length $L$. From these parameters, we define the relative distributed weight $\omega$ such that $\omega = gA(\rho - \rho_f)$.

Therefore, horizontal and vertical displacements are given by Eqn. (18) and (19), horizontal and vertical tensions by Eqn. (20) and (21). Note that in Eqn. (18) and (19), the shape functions $X(s)$ and $Z(s)$ are defined from end point $A$ in the *local* basis such that $s$ is growing along local $\boldsymbol{X}$ axis, as shown in Fig. 3.

$$X(s) = \frac{N_{x_0}}{\omega}\left[\mathrm{asinh}\left(\frac{N_{z_0} + \omega s}{N_{x_0}}\right) - \mathrm{asinh}\left(\frac{N_{z_0}}{N_{x_0}}\right)\right] + \frac{N_{x_0} s}{EA} \quad (18)$$

$$Z(s) = \frac{N_{x_0}}{\omega}\left[\sqrt{1 + \left(\frac{N_{z_0} + \omega s}{N_{x_0}}\right)^2} - \sqrt{1 + \left(\frac{N_{z_0}}{N_{x_0}}\right)^2}\right] + \frac{1}{EA}\left(N_{z_0} s + \frac{\omega s^2}{2}\right) \quad (19)$$

$$N_x(s) = N_{x_0} \quad (20)$$

$$N_z(s) = N_{z_0} + \omega s \quad (21)$$

Because previous fields are defined locally, we may define for convenience the *global* coordinates. In Eqn. (22) to (25), fields $x, z, n_x, n_z$ in lower case define global values while upper case represents local values. Therefore depending on the choice of $s$ convention (I) - $\boldsymbol{X}$ axis coinciding with $\boldsymbol{x}$ axis - or (II) - $\boldsymbol{X}$ and $\boldsymbol{x}$ are opposite - the sign in the global field functions may be adjusted accordingly. The upper sign (+) stands for convention (I), the lower sign (-) for convention (II).

$$x(s) = x_a \pm X(s) \quad (22)$$

$$z(s) = z_a + Z(s) \quad (23)$$

$$n_x(s) = \pm N_x(s) \quad (24)$$

$$n_z(s) = N_z(s) \quad (25)$$

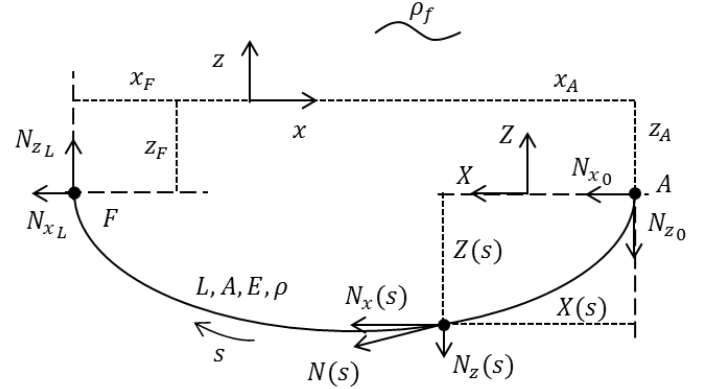

**FIGURE 3. CATENARY CONFIGURATION**

**Semi-analytical resolution process**

The constraints functions associated with boundary conditions (a) to (e) are presented in Table 4. $x_A$ and $z_A$ are the *global* coordinates of the starting point, potentially unknown. $x_F$ and $z_F$ are also the *global* coordinates of the known end point when necessary.

**TABLE 4. BCS CONSTRAINT EQUATIONS**

| Joint type | Constraint equation | |
|---|---|---|
| (a) | $\left.\begin{array}{l}x(s = L) - x_F \\ z(s = L) - z_F\end{array}\right\} = \mathbf{0}$ | (26) |
| (b) | $\left.\begin{array}{l}n_x(s = L) - F_x \\ n_z(s = L) - F_z\end{array}\right\} = \mathbf{0}$ | (27) |
| (c) | $\left.\begin{array}{l}n_x(s = L) - k(a_x - x(s = L)) \\ n_z(s = L) - k(a_z - z(s = L))\end{array}\right\} = \mathbf{0}$ | (28) |
| (d) | $\left.\begin{array}{l}n_x(s = L) - F_x \\ z(s = L) - z_F\end{array}\right\} = \mathbf{0}$ | (29) |
| (e) | $\left.\begin{array}{l}n_x(s = L) - k(a_x - x(s = L)) \\ z(s = L) - z_F\end{array}\right\} = \mathbf{0}$ | (30) |

Constraints equations depend on the initial point positions $x_A$ and $z_A$ as well as initial tensions $n_{x_0}$ and $n_{z_0}$ by means of $x(s)$, $z(s)$, $n_x(s)$ and $n_z(s)$. As mentioned in previous paragraphs, at least two of the four parameters are known depending on the BC considered. Therefore Eqn. (26) through (30) constitute an algebraic 2x2 system $(S)$ depending on two of the four variables $n_{x_0}$, $n_{z_0}$, $x_A$ and $z_A$. A root finding algorithm from Python/Scipy optimize library is used to solve $(S)$. A relative error tolerance of 1.e-12 between two iterations is used in the semi-analytical resolution process. Once $(S)$ solved at BCs, analytical line shape and internal tensions are computed at any curvilinear abscissa through Eqn. (18) to (21).

The semi-analytic method presented above does not differ a lot from the shooting method. It is actually a special case were external forces are imposed to be the weight only. In such restriction, the IVP step of the shooting method can be integrated analytically to obtain Eqn. (18) to (21). Unfortunately for arbitrary external forces, an algebraic closed form cannot be obtained from an analytical integration and one has to use a numerical integration to evaluate the sought fields at $s = L$ in order to calculate the constraints vector. This is the main difference between the above presented shooting method and the "shooting methods" used for catenary solution in commercial software which are closer to the semi-analytical process described in this paper.

**RESULTS**

Even if the semi analytical form of the reference has been developed in the $x - z$ plane, the shooting equations are kept 3D. Numerical simulations are performed with a specifically developed C++ code.

Numerical results are compared to semi-analytical ones, based on the absolute error of field values $\boldsymbol{n}$ and $\boldsymbol{r}$ at both extremities, which are respectively set dimensionless with respect to string weight $wL$ and length $L$. Because both fields $\boldsymbol{n}$ and $\boldsymbol{r}$ will be analytically and numerically (at integration point) available, the maximum dimensionless absolute error along the string will also be computed: internal tension and displacement maximum absolute errors are respectively defined as $\varepsilon_a = \max_{s \in [0,L]} |(n_i - n_{i\ exact})|/wL$ and $\varepsilon_a = \max_{s \in [0,L]} |(x_i - x_{i\ exact})|/L$. The segment properties considered are summarized in Table 5.

**TABLE 5. PHYSICAL PROPERTIES**

| | |
|---|---|
| Length at rest $L$ | 50 m |
| Young Modulus $E$ | 2.11e11 N/m² |
| Cross section area $A$ | 3.1426e-04 m² |
| Material volumic mass $\rho$ | 7.850e03 kg/m³ |
| Fluid volumic mass $\rho_F$ | 1.025e03 kg/m³ |

For case (b) we consider a pure imposed horizontal force, hence $F_z = 0$. For both (b) and (d) we introduce the dimensionless ratio $c = wL/F_x$ such that $F_x$ is obtained for $c = 10$. Moreover left end coordinates for the imposed starting values of convention (I) are $x_A = 0$, $z_A = 0$. Finally right end coordinates for cases (I)-(a), (I)-(d) and (I)-(e) are $x_F = 25$ and $z_F = 0$. The spring reference point position vector is taken as $\boldsymbol{a} = [25, 0, 0]^T$.

For the shooting algorithm, the Newton algorithm tolerance is set up to 1.e-8, and the RKF45 absolute error is taken to be 1.e-8. When starting positions are unknown such as in cases (II)-(b), (II)-(d) and (II)-(e), a first guess is imposed to $\boldsymbol{r}_F = [x_F, z_F]$. In the same way, when initial tension in unknown, i.e. cases (I)-(a-e), (II)-(a), (II)-(d) the vertical and horizontal tensions are supposed equal in magnitude such that the resultant is obtained for $c = 10$. This corresponds to a line starting at 45 degrees downward. Table 7 and Table 8 give the results for the ten configurations considered.

Note that zero absolute errors of the directly imposed field values at the initial end point have been removed from Table 6.

One can see that the dimensionless absolute errors do not exceed 2.92E-09 in $x$ position inside the segment. At initial and final ends, all fields confounded, the maximum error is respectively of 8.29E-10 and 2.18E-09. Obviously these values are extremely dependent on the RKF45 relative error parameter as well as the Newton tolerance selected. Here, the Newton tolerance will mainly influence the error at boundaries whereas the RKF45 relative error will affect the maximum error all along the segment. These results highlight that several boundary conditions types can be used with the shooting method and whatever the boundary conditions configuration chosen, the maximum error of positions and internal load fields are of the same order than errors of unknown fields at boundary conditions. Hence, the present shooting method application being intrinsically continuous without further approximation than (i)-(ii)-(iii), strong accuracy is obtained on geometrical and internal tension fields all along the line as shown in Table 8. Catenary solutions are easily obtained with the present set of shooting equations. Indeed, for distributed loads not depending on the cross section configuration, the IVP is better conditioned leading to a constraint function having a better radius of convergence [4]. However when distributed loads depend on the actual deformed shape (following distributed loads) the single shooting method may have some troubles to converge. In these cases, stabilization techniques such as multi-shooting may be used to improve the problem's conditioning [12,15].

**TABLE 6. INITIAL END POINT ABSOLUTE DIMENSIONLESS ERRORS OF ESTIMATED FIELD**

| Conv. | Case | Fields | | | |
|---|---|---|---|---|---|
| | | $x$ | $z$ | $n_x$ | $n_z$ |
| (I) | (a) | - | - | 2.34E-10 | 6.12E-10 |
| | (b) | - | - | 1.86E-14 | 1.86E-13 |
| | (c) | - | - | 2.90E-10 | 6.43E-10 |
| | (d) | - | - | 1.86E-14 | 6.92E-10 |
| | (e) | - | - | 2.73E-10 | 7.46E-10 |
| (II) | (a) | - | - | 2.34E-10 | 6.12E-10 |
| | (b) | 7.60E-10 | 3.03E-11 | - | - |
| | (c) | 4.10E-11 | 6.80E-12 | 2.39E-10 | 5.72E-10 |
| | (d) | **8.29E-10** | - | 1.86E-14 | 6.92E-10 |
| | (e) | 4.79E-11 | - | 2.73E-10 | 7.46E-10 |

**TABLE 7. FINAL END POINT ABSOLUTE DIMENSIONLESS ERRORS**

| Conv. | Case | Fields | | | |
|---|---|---|---|---|---|
| | | $x$ | $z$ | $n_x$ | $n_z$ |
| (I) | (a) | 2.00E-13 | 4.51E-15 | 2.34E-10 | 6.12E-10 |
| | (b) | **2.18E-09** | 1.13E-10 | 1.86E-14 | 1.90E-15 |
| | (c) | 5.06E-11 | 4.78E-12 | 2.90E-10 | 6.43E-10 |
| | (d) | 8.29E-10 | 4.77E-16 | 1.86E-14 | 6.92E-10 |
| | (e) | 4.79E-11 | 2.53E-15 | 2.73E-10 | 7.46E-10 |
| (II) | (a) | 1.04E-13 | 2.88E-15 | 2.34E-10 | 6.12E-10 |
| | (b) | 1.36E-15 | 2.93E-16 | 1.86E-14 | 1.86E-13 |
| | (c) | 7.97E-16 | 1.06E-16 | 2.39E-10 | 5.72E-10 |
| | (d) | 4.22E-15 | 1.71E-15 | 1.86E-14 | 6.92E-10 |
| | (e) | 1.78E-15 | 6.90E-17 | 2.73E-10 | 7.46E-10 |

**TABLE 8. MAXIMUM SEGMENT FIELDS ABSOLUTE DIMENSIONLESS ERRORS**

| Conv. | Case | Fields | | | |
|---|---|---|---|---|---|
| | | $x$ | $z$ | $n_x$ | $n_z$ |
| (I) | (a) | 8.83E-10 | 1.10E-10 | 2.34E-10 | 6.12E-10 |
| | (b) | **2.92E-09** | 1.59E-10 | 1.86E-14 | 4.42E-13 |
| | (c) | 1.09E-09 | 1.16E-10 | 2.90E-10 | 6.43E-10 |
| | (d) | 1.26E-09 | 1.20E-10 | 1.86E-14 | 6.92E-10 |
| | (e) | 9.47E-10 | 1.13E-10 | 2.73E-10 | 7.46E-10 |
| (II) | (a) | 8.83E-10 | 1.10E-10 | 2.34E-10 | 6.12E-10 |
| | (b) | 1.18E-09 | 6.03E-11 | 1.86E-14 | 1.86E-13 |
| | (c) | 9.54E-10 | 9.66E-11 | 2.39E-10 | 5.72E-10 |
| | (d) | 1.20E-09 | 1.20E-10 | 1.86E-14 | 6.92E-10 |
| | (e) | 9.55E-10 | 1.13E-10 | 2.73E-10 | 7.46E-10 |

**CONCLUSION**

This paper offers an alternative way to model and compute 3D static mooring lines using strings equation, based on the shooting method. Equations solved are fully nonlinear with exact geometrical description without small-displacements hypotheses. With the presented formalism, cross section properties and spatial coordinates as well as internal forces and strains are kept continuous with regards to arc length. All sets of equations, including the contribution of external distributed loads are written in 3D. The constitutive relation is written without assumption on the stretch deformation and the stretch itself is only depending on the internal tension norm. An elastic constitutive law example is introduced and used in the validation cases.

The present work has proposed a string resolution method that keeps a state vector written with regard to both position and force fields. This keeps the approach simple, combined with a low number of degrees of freedom (only three). Hence compared to discretized methods, the computation of the Jacobian inverse is very effective.

Moreover in finite element and finite difference methods, imposed twists and wrenches of the considered mechanical linkage are split into two parts. One constrains the state vector (nodal positions), the other is added to the external nodal forces (second member). This leads to cumbersome manipulation in case the joint's statics depends on the joint's kinematics. In the presented shooting method, imposed twists and wrenches constrain only and directly the state vector of the differential problem.

Boundary conditions written both with respect to spatial coordinates and tension fields allow the possibility of modeling the various Dirichlet, Robin and mixed boundary conditions types in the same way.

The main mechanical linkages used in offshore - spherical, linear annular, free, spring and prismatic/spring - are developed within this formalism. The five proposed configurations and ten validation cases demonstrate the accuracy of the method.

A semi-analytical reference solution based on algebraic catenary closed form equations is given. The reference solution is also adapted to deal with all studied configurations. It is highlighted that the shooting method and this semi-analytic approach are close to each other. Indeed the latter is often used in offshore commercial codes and consists in analytically integrating the shooting method initial value problem because external distributed forces are limited to gravity only, leading to a catenary shape.

Numerical results are in good agreement with the semi-analytical reference in terms of absolute errors all along the line as well as at end points. It shows that the shooting method, as presented, successfully models complex kinematic-spring joints as a generalization of classical connections.

**ACKNOWLEGMENTS**

The authors would like to thank Dr. Gazzola Thomas who provided the spark that led to this work.
**REFERENCES**

[1] Masciola Marco, Jonkman Jason and Robertson Amy. "Extending the capabilities of the mooring analysis program: A survey of dynamic mooring line theories for integration into FAST". *Proceedings of the ASME 2014 33rd International Conference on Ocean, Offshore and Arctic Engineering*. Vol. 9A, OMAE2014-23508: pp. V09AT09A032. San Francisco, CA, June 8-13, 2014. DOI 10.1115/OMAE2014-23508.

[2] De Zoysa, Arjuna P. K. "Steady-state Analysis of Undersea Cables". *Ocean Engineering*. Vol. 5 No. 3 (1978): pp. 209-223.

[3] Friswell, Michael Ian. "Steady-state Analysis of Underwater Cables". *Journal of Waterway, Port, Coastal and Ocean Engineering*. Vol. 121 No. 2 (1995): pp. 98-104.

[4] Surmont, Florian, Coache, Damien, Gazzola, Thomas, Brun, Cédric and Martin, Sébastien. "An Alternative Methodology for Static 3D Nonlinear Hanging Mooring Lines Analysis", *Proceedings of the Twenty-seventh (27th) International Ocean and Polar Engineering Conference*. Vol. 2: pp. 105-112. San Francisco, CA, June 25-30, 2017.

[5] Bailey, Paul and Shampine L. F. "On shooting methods for Two-Point Boundary Value Problems". *Journal of Mathematical Analysis and Applications* Vol. 23 No. 2 (1968): pp. 235-249.

[6] Osborne M. R. "On Shooting Methods for Boundary Value Problems". *Journal of Mathematical Analysis and Applications* Vol. 27 No. 2 (1969): pp. 417-433.

[7] Till, John, Bryson, Caroline, Chung, Scotty and Rucker, Caleb. "Efficient Computation of Multiple Coupled Rod Models for Real-Time Simulation Control of Parallel Continuum Manipulators". *Proceedings of the IEEE International Conference on Robotics and Automation*. pp. 5067-5074. Seattle, WA, May 26-30, 2015. DOI 10.1109/ICRA.2015.7139904.

[8] Vialard, François-Xavier, Risser, Laurent, Rueckert, Daniel and Cotter, Colin J. "Diffeomorphic 3D Image Registration via Geodesic Shooting Using an Efficient Adjoint Calculation". *International Journal of Computer Vision* Vol. 97 No. 2 (2012): pp. 229-241. DOI 10.1007/s11263-011-0481-8.

[9] Sundararajan, P. and Noah, S. T. "Dynamics of Forced Nonlinear Systems Using Shooting/Arc Length Continuation Method – Application to Rotor Systems". *Journal of Vibration and Acoustics* Vol. 119 No. 1 (1997): pp. 9-20.

[10] Keller, Herbert B. *Numerical Methods for Two-Point Boundary Value Problems*. Blaisdell, Waltham, Massachusetts (1968).

[11] Keller, Herbert B. *Numerical Solution of Two-Point Boundary Value Problems*. Vol 24, SIaM (1976).

[12] Stoer, Josef and Bulirsch, Roland. *Introduction to Numerical Analysis*. Springer, New York, Third Edition, (2002).

[13] Deuflhard, Peter. "A Modified Newton Method for the Solution of Ill-Conditioned Systems of Nonlinear Equations with Application to Multiple Shooting". *Numerical Mathematics* Vol. 22 No. 4 (1974): pp. 289-315.

[14] Holsapple, Raymond, Venkataraman, Ram and Doman David. "A Modified Simple Shooting Method for Solving Two-Point Boundary Value Problems". *Proceedings of the IEEE Aerospace Conference*. Vol. 6-2783: pp. 2783-2790. Big Sky, MT, March 8-15, 2003. DOI 10.1109/AERO.2003.1235204.

[15] Liu, Chein-Shan. "The Lie-Group Shooting Method for Solving Multi-dimensional Nonlinear Boundary Value Problems". *Journal of Optimization Theory and Applications* Vol. 152 No. 2 (2012): pp. 468-495. DOI 10.1007/s10957-011-9913-4.

[16] Antman, Stuart Sheldon. *Nonlinear Problems of Elasticity*. Springer-Verlag, New York, Second edition (2005).

[17] Irvine H. Max. *Cable Structures*. Dover Publications, Mineola, New York (1992).

[18] Salençon, Jean. *Mécanique des milieux continus: Milieux curvilignes (Volume 3)*. Editions Ecole Polytechnique, Palaiseau (2016).
9